\documentclass{article}
\usepackage{graphicx}

\begin{document}

\title{\sc Lessons from Schwinger Effective Action for Black Holes}
\author{Sang Pyo Kim\\
{\small  Department of Physics, Kunsan National University, Kunsan 573-701, Korea} \\
{\small {\it Email} : sangkim@kunsan.ac.kr}}

\date{}
\maketitle
\begin{abstract}
We revisit the Hawking radiation by comparing the effective actions in the in-out formalism, and
advance an interpretation of the vacuum polarization and the Hawking radiation. The equivalence
exists between the spinor QED effective action in a constant electric field and the nonperturbative
effective action of a massless boson on the horizon of a Schwarzschild black hole.
\end{abstract}
\section{Introduction}

Pair production in strong background fields has been one of the most important issues in
theoretical physics since the computation of the one-loop effective action in a constant
electromagnetic field by Heisenberg and Euler\,\cite{HeisenbergEuler} and Schwinger\,\cite{Schwinger}
and the discovery of the black hole radiation by Hawking\,\cite{Hawking}. The virtual pairs from vacuum fluctuations
are separated into real pairs by the strong electric field in the Schwinger mechanism and by the causal horizon
of the black hole in the Hawking radiation, as summarized in Table 1.

The pair production is accompanied by the vacuum polarization, that is, the real part of the nonperturbative
effective action. In quantum electrodynamics (QED), for instance, the mean number of
pairs or the vacuum persistence (twice the imaginary
part of the effective action) is closed related to the pole structure of the vacuum polarization.
In the in-out formalism based on the Schwinger variational principle, the effective action is the
scattering matrix amplitude between the in- and the out-vacua, which can be
manifestly realized by the Bogoliubov transformation method\,\cite{DeWitt}.

In this talk, we revisit the new approach to the vacuum polarization and the Hawking radiation of a
Schwarzschild black hole in analogy with the Heisenberg-Euler and Schwinger effective action in QED\,\cite{KimHwang11}.
Though it results from quantum field theory at one-loop, not from quantum gravity, the
nonperturbative effective action, however, may still shed light on quantum aspects of black holes.

\begin{table}
\caption{\label{Tab1} Strong Field Physics: Analogy between QED and Black Hole}
 \begin{tabular}{lll}
  & {\bf Strong QED} & {\bf Black Hole}\\
 \hline
External agent & Electric field & Event horizon \\
Pair production & Schwinger mechanism & Hawking radiation\\
Nonperturbative action & Vacuum polarization  & Stress tensor \\
\hline
 \end{tabular}
\end{table}

\section{Schwinger Mechanism and Effective Action}

The vacuum polarization and the pair production have been systematically studied in spinor QED by Heisenberg and Euler
and in scalar as well as spinor QED by Schwinger. The vacuum polarization may be written as\,\cite{Schwinger}
\begin{eqnarray} \label{QED vac}
{\cal L}_{\rm eff} = (-1)^{2 \sigma} \frac{(1 + 2 \sigma)}{2} \frac{qE}{2 \pi} \int \frac{d^2 {\bf k}_{\perp}}{(2\pi)^2}
\, {\cal P} \int_{0}^{\infty} \frac{ds}{s} \exp \Bigl(- \frac{m^2 + {\bf k}_{\perp}^2}{2qE} s \Bigr)
\nonumber\\ \times \Bigl[\frac{\cos^{2 \sigma} (s/2)}{\sin(s/2)} - \frac{2}{s} + (-1)^{2 \sigma} \frac{1 - \sigma}{6}s \Bigr],
\label{qed act}
\end{eqnarray}
where $\sigma = 0$ for scalar QED and $\sigma = 1/2$ for spinor QED.  The vacuum persistence,
twice the sum of residues at simple poles of the vacuum polarization, is given by
\begin{eqnarray} \label{QED per}
2 {\rm Im} {\cal L}_{\rm eff} = (-1)^{2 \sigma} \frac{(1 + 2 \sigma) (qE)}{2 \pi} \int \frac{d^2 {\bf k}_{\perp}}{(2\pi)^2}
\ln \Bigl( 1 + (-1)^{2 \sigma} {\cal N}_{\bf k} \Bigr), \label{qed per}
\end{eqnarray}
where the mean number of produced pairs and the inverse temperature from the Unruh effect\,\cite{HwangKim09} are
\begin{eqnarray}
{\cal N}_{\bf k} = e^{- \beta (\frac{{\bf k}_{\perp}^2}{2m} + \frac{m}{2})}, \quad \beta = \frac{2 \pi}{(qE/m)}.
\end{eqnarray}
The inversion of spin-statistics has been argued in the vacuum polarization\,\cite{MGR,LabunRafelski} and in the vacuum persistence\,\cite{HwangKim09}, but its physical origin and meaning has not been understood yet.
The Schwinger limit is the critical strength for $e^{-}e^{+}$ pair production,
$E_c = m^2/|e| = 1.3 \times 10^{16} \, {\rm V/cm}$.

In the in-out formalism the Schwinger variational principle leads to the effective action\,\cite{DeWitt}
\begin{eqnarray}
e^{iW} = e^{i \int d^D x \sqrt{-g} {\cal L}_{\rm eff}} = \langle 0, {\rm out} \vert 0, {\rm in} \rangle. \label{W}
\end{eqnarray}
The effective action (\ref{W}) is equivalent to summing the Feynman diagrams in Figure 1.
The pair production necessarily makes the effective action complex since $\vert 0, {\rm out} \rangle \neq
\vert 0, {\rm in} \rangle$. Further, the vacuum persistence and the
mean number of produced pairs are related through
\begin{eqnarray}
e^{- 2 {\rm Im} W} = \vert \langle 0, {\rm out} \vert 0, {\rm in} \rangle \vert^2, \quad
2 {\rm Im} W = (-1)^{2 \sigma} VT \sum_{\bf k} \ln [1 +(-1)^{2 \sigma} {\cal N}_{\bf k}]. \label{vac per}
\end{eqnarray}
In the above $2 {\rm Im} W/(VT) = 2 {\rm Im} {\cal L}_{\rm eff}$ is the decay-rate of the in-vacuum per unit volume and per unit time
and for a small pair-production rate, $2 {\rm Im} {\cal L}_{\rm eff} \simeq \sum_{\bf k} {\cal N}_{\bf k}$.
\begin{figure}[h]
\begin{center}
\includegraphics[width=5.5cm,height=3.5cm]{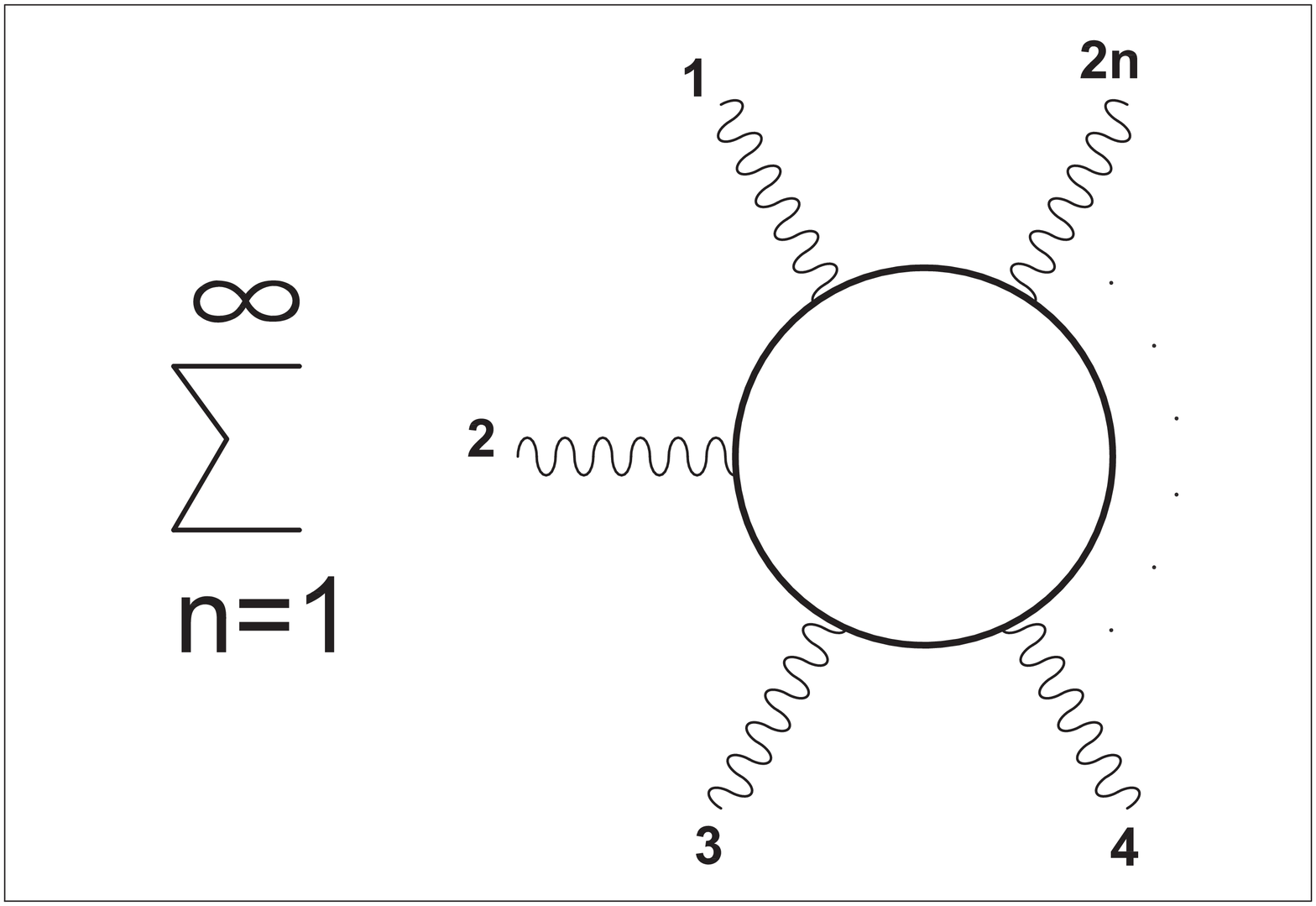}
\caption{One-loop diagrams: the internal loop denotes a charged particle and the external legs (wave lines) denote the
background photons and/or gravitons.}
\label{fig_1}
\end{center}
\end{figure}

Recently Kim, Lee and Yoon have further developed the in-out formalism and introduced the gamma-function
regularization ($\Gamma$-regularization)\,\cite{KLY08,KLY10a,Kim11a}.
The zero-temperature effective action for bosons and fermions is given by
\begin{eqnarray} \label{eff ac}
\frac{W}{VT} = {\cal L}_{\rm eff} = (-1)^{2 \sigma} \sum_{\bf k} \ln \alpha_{\bf k}^*. \label{eff act}
\end{eqnarray}
Here $\alpha_{\bf k}$ is the Bogoliubov coefficient between the out- and the in-vacua for each quantum number ${\bf k}$
\begin{eqnarray}
\hat{a}_{{\rm out}, {\bf k}} = \alpha_{\bf k} \hat{a}_{{\rm in}, {\bf k}} + \beta_{\bf k} \hat{a}^{\dagger}_{{\rm in}, {\bf k}},
\end{eqnarray}
and the coefficients satisfy the relation from the spin-statistics theorem
\begin{eqnarray}
|\alpha_{\bf k}|^2 + (-1)^{2 \sigma} |\beta_{\bf k}|^2 =1.
\end{eqnarray}
The mean number of produced pairs in (\ref{vac per}) is given by ${\cal N}_{\bf k} = |\beta_{\bf k}|^2$.
In a constant electric field, the Bogoliubov coefficient may be found from the spin-diagonal component of the Dirac 
or the Klein-Gordon equation
\begin{eqnarray}
\alpha_{\bf k} = \frac{\sqrt{2 \pi}}{\Gamma (-p)} e^{- i (p \pm 1) \frac{\pi}{2}}, \quad p = - \frac{1}{2} \mp \frac{i}{2 \pi}
{\cal S}_{\bf k},
\end{eqnarray}
where the upper (lower) sign is from the time-dependent (Coulomb) gauge and ${\cal S}_{\bf k} = (m^2 + {\bf k}^2_{\perp} - 2 i \sigma
qE)/(2qE)$ is the instanton action\,\cite{KLY10a}.

Table 2 summarizes the background fields in which the pair production and/or the effective actions have been
known. The in-out formalism has proved a consistent and computationally powerful method for the
effective action and/or the pair production for an electromagnetic field in a curved spacetime such as
de Sitter (dS) space or anti-de Sitter (AdS) space. Since the Bogoliubov coefficients
can be derived from the exact solution of the field equation, it is expected that the effective action
may be found when the background field and/or the spacetime have certain symmetry, leading to
the exact solution. For instance, the Dirac or the Klein-Gordon equation
in a constant electric field has the spectrum generating algebra $SU(1,1)$ and dS and AdS spaces have
the maximal symmetry of the given dimensions.
\begin{table}
\caption{\label{Tab2} Exact Effective Action and/or Pair Production}
 \begin{tabular}{lll}
 {\bf Background Fields} & {\bf EA}
 and {\bf PP} & {\bf Reference}\\
 \hline
 Constant EM-field & EA and PP & Heisenberg-Euler\,\cite{HeisenbergEuler} \\
 & & Schwinger\,\cite{Schwinger} \\
Sauter-type E-field  & PP & Nikishov\,\cite{Nikishov} \\
Sauter-type E-field  & EA and PP & Dunne-Hall\,\cite{DunneHall}\\
& & Kim-Lee-Yoon\,\cite{KLY08,KLY10a}\\
E-field in dS and AdS space & PP & Kim-Page\,\cite{KimPage08}\\
& & Kim-Hwang-Wang\,\cite{KHW}\\
dS space & EA and PP & Kim\,\cite{Kim10-dS}\\
\hline
EA: effective action & PP: pair production &
 \end{tabular}
\end{table}

\section{Vacuum Polarization and Hawking Radiation}

The Hawking radiation of bosons and fermions from a charged rotating black hole is given by\,\cite{Page}
\begin{eqnarray}
N_{J} (\omega) = \frac{1 - |R_J|^2}{ e^{\beta (\omega - m \Omega_H - q \Phi_H)} + (-1)^{2 \sigma }}, \quad \beta = \frac{1}{k_{\rm B} T_{\rm H}}, \quad
T_{\rm H} = \frac{\kappa}{2 \pi}.
\end{eqnarray}
Here $R_J$ is the amplification factor, $\Omega_H$ the angular momentum of the hole,
 $\Phi_H$ the electric potential and $\kappa$ the surface gravity
on the event horizon. In the case of the zero amplification factor, the
vacuum persistence is
\begin{eqnarray}
2 {\rm Im} W = - (-1)^{2 \sigma } \sum_J \ln (1 -(-1)^{2 \sigma} e^{- \beta (\omega - m \Omega_H - q \Phi_H) } ). \label{bh per}
\end{eqnarray}
Note the change of sign in contrary to the QED case.

A four-dimensional Schwarzschild black hole with mass $M$ has the inverse temperature $\beta = 8 \pi M$.
Denoting $J = \{\omega, l, m, p \}$, with the spherical harmonics $l, m$ and
the polarization $p$ and the energy $\omega$, the Bogoliubov coefficients for a massless boson field are found\,\cite{DeWitt}
\begin{eqnarray}
\alpha_{J} = A_{J} e^{2 \pi M \omega} \Gamma (1 + i 4M \omega), \quad
\beta_{J} = - A_{J} e^{-2 \pi M \omega} \Gamma (1 + i 4M \omega).
\end{eqnarray}
Now the effective action (\ref{eff act}) takes the form
\begin{eqnarray}
W = i (8 \pi M) \sum_{l} (2l+1) (2p+1) \int \frac{d \omega}{2 \pi} \ln \Gamma(1 - i 4M \omega).
\end{eqnarray}
Employing the $\Gamma$-regularization, we find the effective action per unit horizon area\,\cite{KimHwang11}
\begin{eqnarray}
{\cal L}_{\rm eff} = - \frac{1}{16 \pi M} \sum_{l} (2l+1) (2p+1) \int \frac{d \omega}{2 \pi}\,
{\cal P} \int_{0}^{\infty} \frac{ds}{s} e^{- 4M \omega s} \Bigl[\frac{\cos(s/2)}{\sin(s/2)} - \frac{2}{s} \Bigr]. \label{bh act}
\end{eqnarray}
It is remarkable that the effective action (\ref{bh act}) and the vacuum persistence (\ref{bh per}) have the form
(\ref{qed act}) and (\ref{qed per}) of spinor QED in a constant electric field.

The vacuum persistence quantifies the decay rate of the vacuum due to the Schwinger mechanism or the Hawking radiation.
Further, it is known that the trace anomalies explain the vacuum persistence, that is, the Schwinger mechanism and
the Hawking radiation. In fact, the vacuum persistence for bosons per unit horizon area\,\cite{KimHwang11}
\begin{eqnarray}
2 {\rm Im} {\cal L}_{\rm eff} = \sum_{l} (2l+1) (2p+1) \frac{\pi}{12} \frac{1}{\beta^2},
\end{eqnarray}
is equal to the total flux from the gravitational anomalies\,\cite{RobinsonWilczek}.

\section{Conclusion}

We have presented the one-loop effective action for QED in a constant electric field
and the Hawking radiation of a Schwarzchild black hole in the in-out formalism.
It consists of the vacuum polarization and the vacuum persistence responsible for pair production.
The prominent feature of the nonperturbative effective action for a Schwarzschild black hole is
that it shares many features in common with spinor QED effective action in a constant electric field.

There remain a few questions to be further pursued: firstly, to find the local
effective action outside the horizon, secondly, to investigate the amplification (grey body) factor,
and thirdly, to find the effective action at two-loop and higher loops.
Still another interesting question is the Schwinger effect in a Reissner-N\"{o}rstrom black hole.
Finally, the origin of spin-statistics inversion of QED differently from gravity challenges a
further study\,\cite{KimHwang11,HwangKim09,MGR,LabunRafelski}.

\section{Acknowledgements}
The author would like to thank W-Y.~Pauchy Hwang, Hyun Kyu Lee and Yongsung Yoon for early collaborations and
Eun Ju Kang for drawing the figure.
The participation of ICGAC10 was supported by Basic Science Research Program through
the National Research Foundation of Korea (NRF) funded by the Ministry of Education, Science and Technology (2011-0002-520).
The work of this paper was supported in part by National Science Council Grant (NSC 100-2811-M-002-012), Taiwan.


\begin{thebibliography}{00}

\bibitem{HeisenbergEuler} W.~Heisenberg and H.~Euler, {\em Folgerungen aus der Diracschen
Theorie des Positrons\/}, Z. Phys. 98 (1936) 714.

\bibitem{Schwinger} J.~Schwinger, {\em On gauge invariance and vacuum polarization\/}, Phys.\ Rev.\ 82 (1951) 664.

\bibitem{Hawking} S.~W.~Hawking, {\em Particle creation by black holes\/}, Commun.\ Math.\ Phys. 43 (1975) 199.

\bibitem{DeWitt} B.~S.~DeWitt, {\em Quantum Field Theory in Curved Spacetime\/}, Phys.\ Rept.\  19 (1975) 295.

\bibitem{KimHwang11} S.~P.~Kim and W-Y.~P.~Hwang, {\em Vacuum Polarization and Persistence on the Black Hole Horizon\/}
[arXiv:1103.5264].

\bibitem{HwangKim09} W-Y.~P.~Hwang and S.~P.~Kim, {\em Vacuum persistence and inversion of
spin statistics in strong QED\/}, Phys.\ Rev.\ D 80 (2009) 065004.

\bibitem{MGR} B.~M\"{u}ller, W.~Greiner and J.~Rafelski, {\em Interpretation of external fields as temperature\/}, Phys.\ Lett.\ A 63 (1977) 181.

\bibitem{LabunRafelski} L.~Labun and J.~Rafelski, {\em Acceleration and Vacuum Temperature\/}, [arXiv:1203.6148].

\bibitem{KLY08} S.~P.~Kim, H.~K.~Lee and Y.~Yoon, {\em Effective Action of Scalar QED in Electric Field Backgrounds\/},
Phys.\ Rev.\ D 78 (2008) 105013.

\bibitem{KLY10a} S.~P.~Kim, H.~K.~Lee and Y.~Yoon, {\em Action of QED in Electric Field Backgrounds II. Spatially Localized Fields\/},
Phys.\ Rev.\ D 82 (2010) 025015.

\bibitem{Kim11a} S.~P.~Kim, {\em Probing the Vacuum Structure of Spacetime\/} [arXiv:1102.4154].

\bibitem{Nikishov} A.~I.~Nikishov, {\em Barrier scattering in field theory removal of Klein paradox\/}, Nucl.\ Phys.\ B
21 (1970) 346.

\bibitem{DunneHall} G.~V.~Dunne and T.~M.~Hall, {\em QED effective action in time dependent electric
backgrounds\/}, Phys.\ Rev.\ D  58 (1998) 105022.

\bibitem{KimPage08} S.~P.~Kim and D.~N.~Page, {\em Schwinger pair production in dS$_2$ and AdS$_2$\/}, Phys.\ Rev.\ D 78 (2008) 103517.

\bibitem{KHW} S.~P.~Kim, W-Y.~Pauchy Hwang and T-C.~ Wang, {\em Schwinger mechanism in dS$_2$ and AdS$_2$ revisited\/} [arXiv:1112.0885].

\bibitem{Kim10-dS} S.~P.~Kim, {\em Vacuum Structure of de Sitter Space\/} [arXiv:1008.0577].

\bibitem{Page} D.~N.~Page, {\em Hawking radiation and black hole thermodynamics\/},
 New J. Phys. 7 (2005) 203.

\bibitem{RobinsonWilczek} S.~P.~Robinson and F.~Wilczek, {\em Relationship between Hawking Radiation and Gravitational Anomalies\/},
Phys. Rev. Lett. 95 (2005) 011303.
\end{thebibliography}
\end{document}